\documentclass[aps,prl,reprint,superscriptaddress,floatfix,longbibliography]{revtex4-1}

\usepackage[pdftex]{graphicx}
\usepackage[utf8]{inputenc}    % change to your encoding (e.g. utf8)
\usepackage[T1]{fontenc}
\usepackage{amssymb,amsmath}
\usepackage[colorlinks=true,linkcolor=blue,urlcolor=blue,citecolor=blue]{hyperref}

\begin{document}

\title{Measurement-induced macroscopic superposition states in cavity optomechanics}
\date{\today}
\author{Ulrich B. Hoff}
\email{ulrich.hoff@fysik.dtu.dk}
\affiliation{Department of Physics, Technical University of Denmark, Building 309, 2800 Kgs. Lyngby, Denmark}
\affiliation{Australian Research Council Centre of Excellence for Engineered Quantum Systems (EQuS), School of Mathematics and Physics, The University of Queensland, St. Lucia, QLD 4072, Australia}
\author{Johann Kollath-B\"onig}
\author{Jonas S. Neergaard-Nielsen}
\author{Ulrik L. Andersen}
\affiliation{Department of Physics, Technical University of Denmark, Building 309, 2800 Kgs. Lyngby, Denmark}

\begin{abstract}
We present a novel proposal for generating quantum superpositions of macroscopically distinct states of a bulk mechanical oscillator, compatible with existing optomechanical devices operating in the readily achievable bad-cavity limit. The scheme is based on a pulsed cavity optomechanical quantum non-demolition (QND) interaction, driven by displaced non-Gaussian states, and measurement-induced feedback, avoiding the need for strong single-photon optomechanical coupling. Furthermore, we show that single-quadrature cooling of the mechanical oscillator is sufficient for efficient state preparation, and we outline a three-pulse protocol comprising a sequence of QND interactions for squeezing-enhanced cooling, state preparation, and tomography.
\end{abstract}

\maketitle

\paragraph{Introduction.}
Elusive as they are, Schr\"odinger cat~\cite{Schrodinger1935} states remain some of the hardest to tame in the quantum world, yet also among the ones most strived for. That is due to their quintessential embodiment of the manifestly non-classical properties of quantum mechanics, by simultaneously occupying two macroscopically distinct states -- dead \textit{and} alive. Successful creation of such coherent state superpositions have so far been limited exclusively to isolated microscopic quantum systems, e.\,g.~in ion traps~\cite{Monroe1996,Leibfried2005} and microwave cavity and circuit quantum electrodynamics~\cite{Brune1996,Deleglise2008,Kirchmair2013,Vlastakis2013}, while closely related variants, colloquially termed Schr\"odinger kittens, have been demonstrated in propagating optical fields~\cite{Neergaard-Nielsen2006,Ourjoumtsev2007a,Huang2015}. However, an intriguing and long standing question is whether also macroscopic objects can be prepared in quantum superpositions of being here \textit{and} there? 

Within the last decades the field of optomechanics~\cite{Aspelmeyer2014} has attracted tremendous theoretical and experimental attention and has developed into a mature research discipline already contributing a range of key results in quantum physics: mechanical ground state cooling~\cite{Teufel2011,Chan2011}, observation of quantum back-action~\cite{Murch2008,Purdy2013a}, ponderomotive squeezing~\cite{Brooks2012,Safavi-Naeini2013,Purdy2013b}, and recently generation of non-classical mechanical states of motion~\cite{Pirkkalainen2015,Wollman2015,Lecocq2015}. The demonstrated technological ability to engineer mechanical oscillators ranging from micro- to macroscopic in size and tailor their interaction with radiation fields places optomechanics among the most promising testbeds for experimental scrutiny of the long debated quantum-classical transition. Experiments of this kind are of utmost importance for understanding the foundations of quantum mechanics and quantum measurement theory~\cite{Wheeler1983}, and from a technological point of view, engineering and coherent manipulation of mechanical quantum states can be of great use for quantum information processing protocols~\cite{Andersen2015}.  

A vast number of proposals for optomechanical generation of non-Gaussian mechanical states, such as cat states, exist in the literature~\cite{Mancini1997,Bose1997,Akram2013,Ghobadi2014,Paternostro2011,Vanner2013a}. Non-Gaussian states of light can be directly mapped onto the mechanical motional states either via a swapping operation~\cite{Wang2012a,Filip2015} or by teleportation~\cite{Romero-Isart2011,Hofer2011}, but this can be achieved only in the highly challenging sideband resolved regime in which the mechanical frequency lies outside the resonance of a narrow-banded cavity (also known as the good cavity limit). Mechanical non-Gaussian states can also be generated in the much simpler bad cavity regime (where the sidebands are unresolved) by using a broadband cavity and either single photon~\cite{Marshall2003} or coherent state resources~\cite{Carlisle2015}. However, these protocols rely on an extremely strong non-Gaussian interaction between light and mechanics and are thus of limited practical feasibility due to the insufficient optomechanical interaction strengths currently achievable.
One way of bridging the gap is to apply a displacement operation~\cite{Lvovsky2002a} to the optical input state, e.g. a single photon~\cite{Khalili2010,Sekatski2014}. However, whereas this approach enhances the interaction and offers mechanical superposition states of distinguishable constituents in phase space, only a modest degree of macroscopicity (separation in phase space) can be achieved. 

Here, we propose a novel scheme for mechanical cat-like state generation, employing displaced photon subtracted squeezed vacuum (PSSV) states~\cite{Dakna1997} in conjunction with a pulsed measurement induced optomechanical QND interaction~\cite{Vanner2011,Vanner2013}. The proposed scheme relies on the easily accessible sideband unresolved regime, the required optomechanical coupling strength is weak, and the resulting phase space separation of the constituent cat state components is large. For completeness, we also suggest a full three-pulse protocol for pre-cooling, non-Gaussian state preparation and read-out. We find that by using experimentally feasible system parameters, a superposition state of a massive system with a large degree of macroscopicity can be formed.

\begin{figure}[htbp!]
\includegraphics[width=0.85\columnwidth]{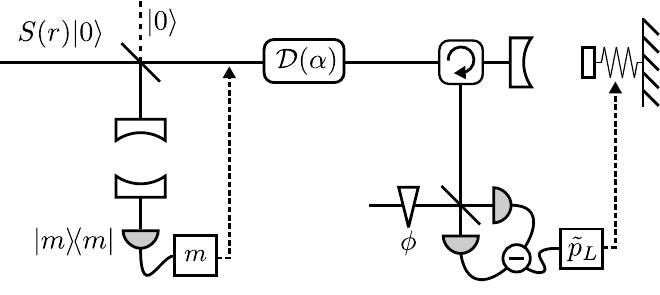} 
\caption{
\label{fig: schematic_representation} 
Employing a displaced PSSV state of light as resource for driving a cavity optomechanical QND interaction, the mechanical oscillator is projected into a highly non-classical state conditioned on the outcome of a subsequent homodyne detection of the optical phase quadrature.}
\end{figure}

The core of the scheme is presented in Fig.~\ref{fig: schematic_representation}. The optical input mode is conditionally prepared in a pulsed PSSV state, by photon number resolved detection on a filtered tap-off of a vacuum squeezed state. Before injection into a cavity optomechanical system, pre-cooled close to its motional ground state, a displacement operation is applied to the optical input to enhance the optomechanical interaction strength. This can be implemented by admixing a strong coherent field on an asymmetric beam splitter. For interaction times much shorter than the mechanical evolution time, $\tau \ll \omega_M^{-1}$, an effective optomechanical QND interaction is realized, and grace to that the mechanics can be projected into a highly non-classical quantum state conditioned on a subsequent measurement on the reflected optical field. In the following, we model the state preparation analytically using the Wigner function formalism. 

\paragraph{Optomechanical interaction.} 
We consider a single-ended cavity optomechanical system excited by an optical pulse of $N_p$ photons and duration $\tau$, much shorter than the free evolution timescale of the mechanical oscillator at frequency $\omega_M$. Furthermore, the cavity bandwidth $\kappa$ (HWHM) is assumed to be much broader than that of the optical pulse. Under these conditions, the dynamics of the optical intra-cavity field can be adiabatically eliminated and mechanical damping and noise processes can be neglected during the interaction time~\cite{Vanner2011}. 
The optical input mode consists of two parts: a quantum fluctuation part described by bosonic operators $a(t),a^{\dagger}(t)$ and with temporal mode function $f(t)$, defined by the conditional PSSV preparation scheme~\cite{Molmer2006} and satisfying $\int dt|f(t)|^2 = 1$, and a classical driving field with amplitude $|\alpha(t)|=\sqrt{N_p}f(t)$ matched to the quantum state. Instantaneous quadrature operators are defined as ${x_L(t) = (a(t)+a^{\dagger}(t))/\sqrt{2}}$ and ${p_L(t) = -i(a(t)-a^{\dagger}(t))/\sqrt{2}}$ with commutator ${[x_L(t),p_L(t')]=i\delta(t-t')}$. To encompass the temporal correlations within the input quantum state, collective quadratures are defined for the entire pulse as $x_L = \int_{-\infty}^{\infty} dt f(t) x_L(t)$ and $p_L = \int_{-\infty}^{\infty} dt f(t) p_L(t)$, obeying $[x_L,p_L] = i$. The mechanical oscillator is similarly described by bosonic operators $b, b^{\dagger}$ and quadrature operators $x_M = (b + b^{\dagger})/\sqrt{2}$, $p_M = -i(b - b^{\dagger})/\sqrt{2}$ with $[x_M,p_M] = i$. Assuming resonant interaction, the linearized Hamiltonian takes the form ${H = -\hbar g_0(2|\alpha(t)| x_L(t) x_M + \sqrt{2} |\alpha(t)|^2 x_M)}$, where the optomechanical coupling constant is given by ${g_0 = x_{zpf}\omega_0/L}$ for an optical Fabry-Perot cavity with length $L$, and $x_{zpf} = \sqrt{\hbar/2M\omega_M}$ is the zero-point fluctuation amplitude of the mechanical oscillator with mass $M$. Integrating the corresponding quantum Langevin equations over the entire interaction results in the following input-output transformation of the quadratures
\begin{subequations}
\begin{align}
x_L^{in} &\rightarrow x_L = x_L^{in} \label{eq: xLout}\\
p_L^{in} &\rightarrow p_L = p_L^{in} + \chi x_M^{in} \label{eq: pLout}\\
x_M^{in} &\rightarrow x_M = x_M^{in} \label{eq: xMout}\\
p_M^{in} &\rightarrow p_M = p_M^{in} + \chi x_L^{in} + \Omega \label{eq: pMout},
\end{align}
\end{subequations}
which is of the well-known QND form~\cite{Grangier1998}. In Eqs.~(\ref{eq: pLout}) and (\ref{eq: pMout}) we have introduced the coupling strength ${\chi = 4g_0\sqrt{N_p}/\kappa}$, weighting the contribution of the measured mechanical position $x_M^{in}$ to the optical output phase quadrature, and the associated back-action momentum transfer $\Omega = \chi \sqrt{N_p/2}$ imposed on the mechanical phase quadrature.

We assume that the initial optical and mechanical states are prepared independently and described by Wigner functions $W^{in}_L$ and $W^{in}_M$, respectively. Through the optomechanical interaction the system degrees of freedom become correlated, the degree of correlation being proportional to the QND coupling strength $\chi$, as evident from Eqs.~(\ref{eq: xLout}) and (\ref{eq: pMout}). By performing a post interaction homodyne measurement of the reflected optical phase quadrature, the mechanical oscillator is projected into an output state $W_M^{out}$, the actual state prepared being conditioned on the measurement outcome $\tilde{p}_L$. Tracing over the unmeasured optical quadrature the conditionally prepared mechanical state is, up to a normalization factor, given by
\begin{align}
W_M^{out}(x_M,p_M) &= \int \!\!\! \int \!dx_L dp_L W_L^{in}(x_L, p_L-\chi x_M)\nonumber\\ 
&\times W_M^{in}(x_M,p_M-\chi x_L - \Omega) \delta(p_L - \tilde{p}_L).
\label{eq: Wmout_general}
\end{align}
Consequently, by driving the optomechanical interaction with a non-classical optical input, the distinctly quantum properties can be transferred to the mechanical oscillator through measurement induced quantum correlations. 

\paragraph{PSSV state preparation.}
Having detailed the optomechanical interaction we now discuss the employed non-classical optical resource state. Assume that a Gaussian state with quadrature variances $V_{x_L}, V_{p_L}$ impinges on a beam splitter with transmittivity $\sqrt{T}$ close to unity (Fig.~\ref{fig: schematic_representation}). For a squeezed vacuum state with squeezing parameter $r$, the $x$ and $p$ variances are $e^{2r}/2$ and $e^{-2r}/2$, respectively.
The reflected field is filtered by a cavity and directed onto a photon number resolving detector, yielding a measurement outcome $m$. Using the Wigner picture beam splitter transformation~\cite{Andersen2009} and tracing out the detected output mode, the resulting conditional PSSV state is given by
\begin{align}
W_L^{in}(x_L,p_L) &= \mathcal{N}\int \!\!\! \int dx_L' dp_L' W_S\left( ax_L-bx_L',ap_L-bp_L' \right) \nonumber\\
& \quad \times W_V\left( ax_L+bx_L',ap_L+bp_L' \right) \nonumber\\
& \quad \times W_D(x_L',p_L').
\end{align}
Here $a=\sqrt{T}$ and $b=\sqrt{1-T}$, and $W_S$ and $W_V$ describe the input phase squeezed vacuum state and the vacuum field admixed in the displacement operation, respectively. $W_D$ is the corresponding Wigner function for the employed optical detection process, which in the following will be taken to be an $m$-photon detector with efficiency $\eta$, $W_D(x, p) = \frac{1}{\pi} \frac{(-\eta)^m}{(2-\eta)^{1+m}} L_m \left( \frac{2x^2 + 2p^2}{2-\eta} \right) \exp \left( -\frac{\eta}{2-\eta} (x^2+p^2) \right)$~\cite{Ferraro2005}.

\paragraph{Conditional mechanical state}
In order for the transferred quantum state to dominate the initial mechanical thermal noise, thereby allowing generation of mechanical states with distinctly non-classical properties, pre-cooling of the mechanical motion is required. The oscillator is initially assumed in thermal equilibrium with a cryogenic bath, and immediately preceding the optomechanical QND interaction, the mechanical mode is cooled close to its motional ground state. Operating in the unresolved sideband regime excludes ordinary sideband laser cooling. However, optical feedback cooling holds promises for reaching the ground state in the bad-cavity limit as well~\cite{Genes2008}, though it remains to be demonstrated~\cite{Wilson2015}. Alternatively, single quadrature cooling-by-measurement~\cite{Vanner2013} can be invoked as discussed later. To accommodate this we describe the pre-cooled mechanical state by a thermal state Wigner function $W_M^{in}$ with distinct quadrature variances $V_{x_M}$ and $V_{p_M}$.
Using Eq.~(\ref{eq: Wmout_general}) together with the PSSV Wigner function $W_L^{in}$ for  the optical input state and $W_M^{in}$, we derive the following analytical expression for the conditional mechanical output state: 
\begin{widetext}
\begin{align}
W_M^{out}(x_M,p_M) &= \mathcal{N}_M \exp \left[ -\frac{1+2V_{x_M}'}{2V_{x_M}}\left(x_M -\frac{2V_{x_M}'}{1+2V_{x_M}'}\frac{\tilde{p}_L}{\chi} \right)^2 
\right]
\exp \left[-\frac{(p_M-\Omega)^2}{(V_{p_M}/V_{p_M}')(1+2V_{p_M}')} \right]\nonumber \\
& \quad\times \sum_{k=0}^{m}\left(\frac{-2}{2-\eta}\right)^k
\begin{pmatrix} m \\ k \end{pmatrix} \sum_{l=0}^k \frac{1}{s_x^{k-l}s_p^l}\left( \frac{2V_{p_M}''+1}{2V_{p_M}'+1}\right)\nonumber \\
& \quad \times L_{k-l}^{-1/2}\left( -\chi^2c_x \left( x_M-\tilde{p}_L/\chi \right)^2 \right) L_l^{-1/2} \left( -\frac{c_p/\chi^2(p_M-\Omega)^2}{(2V_{p_M}''+1)(2V_{p_M}'+1)} \right).
\label{eq: WmechOut}
\end{align}
\end{widetext}
Here, $\mathcal{N}_M$ is a normalization constant and $L_n^m(x)$ the associated Laguerre polynomials. For simplicity, the following lumped parameters have been introduced: 
\begin{subequations}
\begin{align}
s_{x(p)} &= T+2(1-T)V_{x(p)_L} +\eta/(2-\eta)\\
c_{x(p)} &=T(1-T)(1-2V_{x(p)_L})^2/s_{x(p)}\\
V_{x_M}^{\chi} &= \chi^2 V_{x_M} \\ 
V_{p_M}^{\chi} &= V_{p_M}/\chi^2\\
V_{x(p)_M}' &= V_{x(p)_M}^{\chi}\frac{2V_{x(p)_L} + \frac{\eta}{2-\eta} (1 - T + 2TV_{x(p)_L})} {s_{x(p)}}\\
V_{x(p)_M}'' &= V_{x(p)_M}^{\chi} (1-T + 2T V_{x(p)_L})
\end{align}
\end{subequations}
In the following, we will arbitrarily assume an outcome $\tilde{p}_L=0$ of the post-QND optical homodyne measurement. This either corresponds to probabilistic heralding based on the measurement outcome or deterministic real-time feedback actuation of the mechanical oscillator.

\begin{figure*}[!ht]
\includegraphics[width=2\columnwidth]{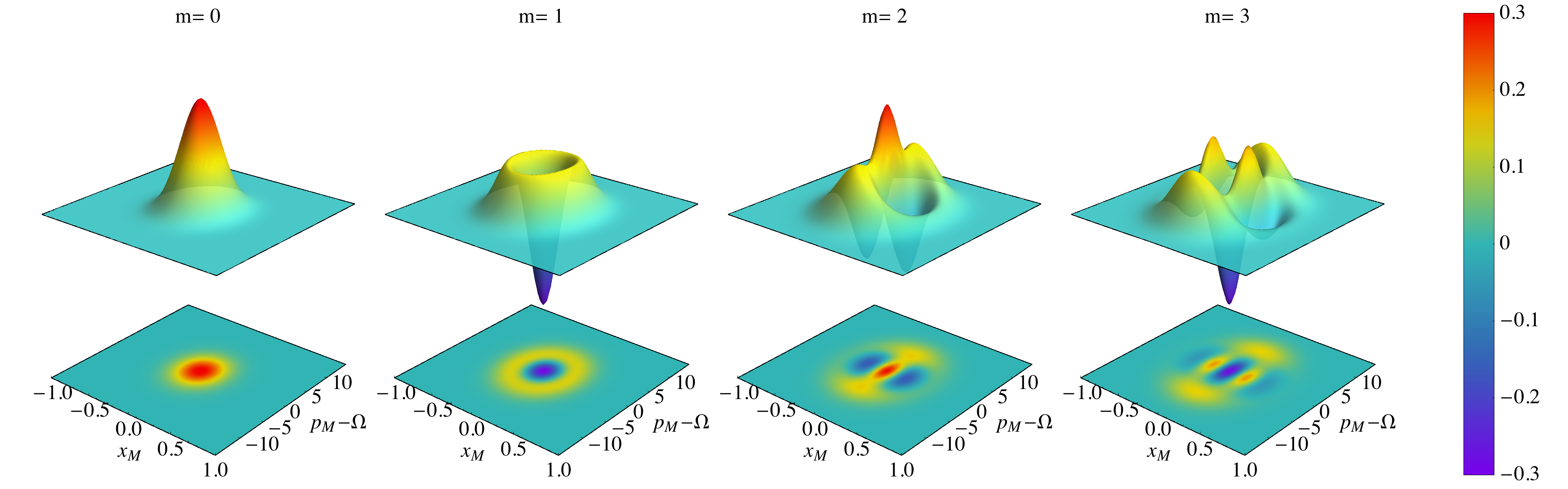}
\caption{Wigner function representation $W_M^{out}$ of the conditionally prepared mechanical states for optical PSSV input states with $m=0$ to $m=3$ photons subtracted from a phase squeezed vacuum state with  $r=1.5$. The QND interaction strength is set to $\chi = 1$ with a corresponding momentum kick of $\Omega = 400$. A tap-off beam splitter transmittivity of $T = 0.98$ and detection efficiency of $\eta=0.95$ is assumed, and the mechanical mode is initially cooled to $\bar{n}=1$.
\label{fig: CondmechWignerstates} 
}
\end{figure*}

\paragraph{Practical feasibility.}
As a feasible system for implementation of the protocol, we propose an optomechanical device merging existing technologies from fiber micro-cavities~\cite{Flowers-Jacobs2012} with tethered ${\rm Si}_3{\rm N}_4$ membrane mechanical resonators~\cite{Norte2015}. In this way the pulsed QND condition $\omega_M \ll \tau^{-1} \ll \kappa$ can be fulfilled while maintaining an apppreciable $g_0/\kappa$ ratio by combining the low frequency and low dissipation mechanical trampoline mode with the small size of the fiber cavity. In particular, we consider an optomechanical Fabry-Perot resonator at $\lambda_L=1550\,\textrm{nm}$ consisting of a vibrating plane end mirror separated $4\,\mu\textrm{m}$ from a concave mirror formed directly at the facet of a fiber. The cavity end mirror is formed by a photonic crystal structure patterned on a tethered membrane with mechanical frequency $\omega_M/2\pi = 100\,\textrm{kHz}$, effective mass $M=1\,\textrm{ng}$, and quality factor $Q_M = 10^8$, and we assume an optical cavity linewidth of $\kappa/2\pi = 1\,\textrm{GHz}$ (${\mathcal{F} \approx 19000}$). The amplitude of the mechanical zero-point fluctuation is $x_{zpf} = 9.1\,\textrm{fm}$ and the resulting optomechanical single photon coupling rate is $g_0/2\pi = 442\,\textrm{kHz}$. Consequently, a QND interaction strength of $\chi=1$ can be achieved using only $N_p = 3.2 \cdot 10^5$ photons in the input pulse. 
The resulting mechanical state Wigner function $W_M^{out}$ is plotted in Fig.~\ref{fig: CondmechWignerstates}, and we observe that as the number of subtracted photons $m$ is progressively increased  the mechanical state indeed approaches that of a Schr\"odinger cat. 
Furthermore, the mechanical state is largely displaced along the phase quadrature as a consequence of the back-action momentum kick $\Omega$. Finally, the strong squeezing of the mechanical mode resulting from the QND interaction should be noted, somewhat masked by the large aspect ratio of the above phase space plots. 

\begin{figure}[tp!]
\includegraphics[width=1\columnwidth]{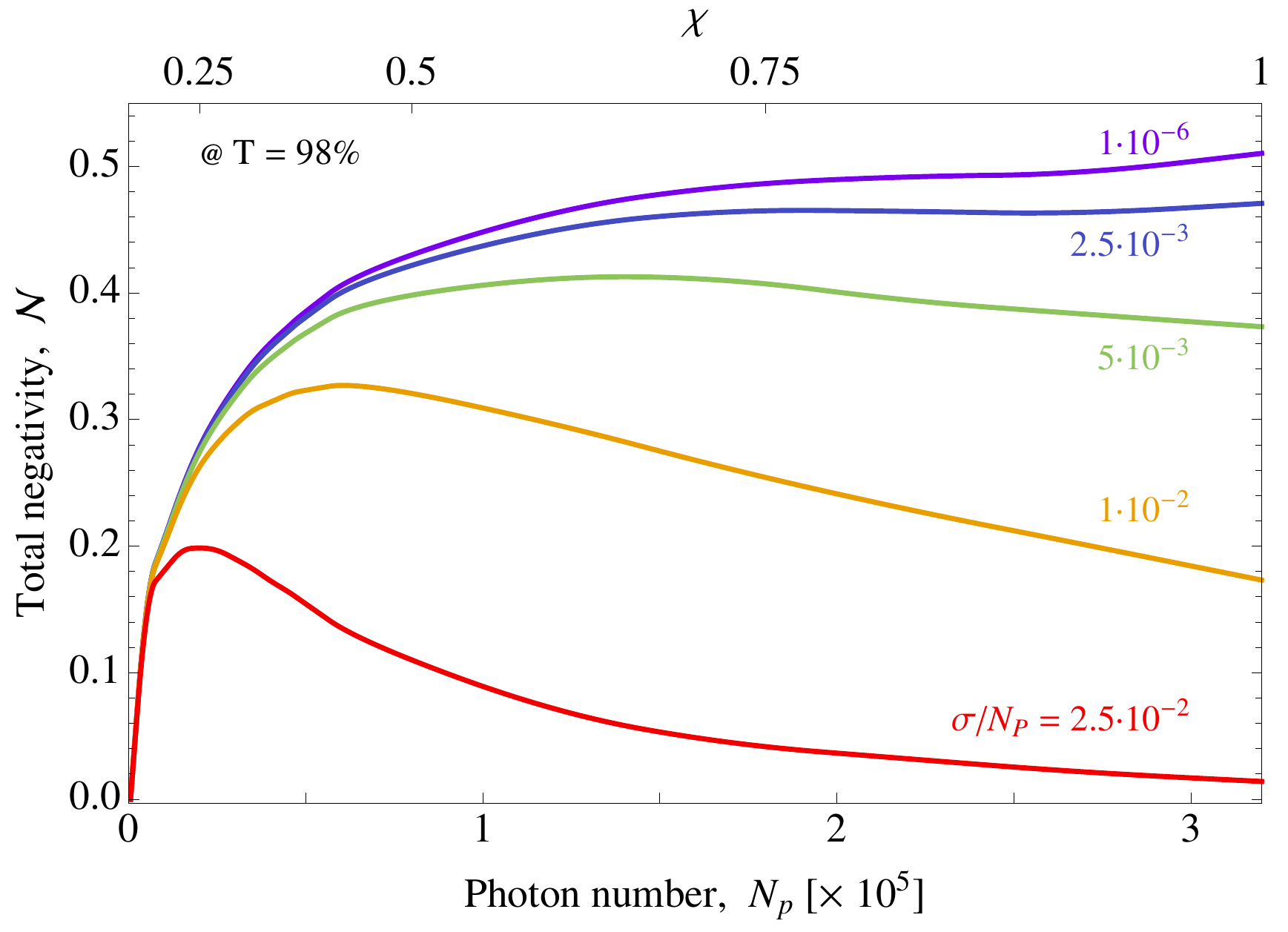}
\caption{Impact of laser amplitude fluctuations on the total negativity of the mechanical state as a function of the employed mean photon number (QND interaction strength) and amplitude fluctuation level. The initial thermal occupancy of the mechanical oscillator mode is set to $\bar{n}=1$, and for the optical input state, $r=1.5$, $m=3$, and a detection efficiency of $\eta=0.95$ is assumed. 
\label{fig: AmplitudeNoise} 
}
\end{figure}

Stability of the phase space displacement is of high importance as fluctuations will smear out the fine structures of the cat state. These fluctuations are amplified by the magnitude of the displacement and it is thus important to keep the displacement at a minimum. In the following, we investigate the effect of amplitude fluctuations on the prepared mechanical state in order to put an upper bound on the permissible fluctuation level. Assuming that the laser amplitude fluctuations obey a Gaussian distribution $\mathcal{G}_{\alpha}(N_p,\sigma_N)$ characterized by mean $N_p$ and standard deviation $\sigma_N$, we model the impact on the mechanical state by a weighted averaging of the mechanical Wigner function
\begin{equation}
\langle W_M^{out}(x_M,p_M) \rangle = \int dN_p W_M^{out}(x_M,p_M; N_p) \mathcal{G}_{\alpha}(N_p,\sigma_N).
\end{equation}
The resulting loss of fringe visibility and thereby quantum coherence is reflected in a degradation of the negativity of the Wigner function. To quantify the effect we therefore use the total negativity~\cite{Kenfack2004,Kleckner2008}
\begin{equation}
\mathcal{N} = \int_{-\infty}^{\infty}\int_{-\infty}^{\infty} dx_M dp_M |W_M^{out}(x_M,p_M)|-1
\end{equation}
as a measure for the degree of preserved non-classicality in the oscillator state. The simulation results presented in Fig.~\ref{fig: AmplitudeNoise} show that the preparation process is highly susceptible to amplitude noise, despite the modest number of photons required to achieve a strong QND interaction in the considered system. A negligible degradation of the total negativity requires $\sigma_N/N_p \leq 10^{-3}$, and for relative fluctuations on the percent level a significant reduction of the maximum total negativity is expected. As the amplitude fluctuations are increased the maximum total negativity occurs for increasingly smaller photon numbers, signifying an optimal trade off between optomechanical coupling strength and loss of quantum coherence.  

\paragraph{Macroscopicity}
The macroscopicity of the prepared mechanical state is assessed by employing the measure proposed by Lee and Jeong~\cite{Lee2011}, which when transformed to $(x,p)$ phase space takes the form
\begin{equation}
\mathcal{I} = -\frac{\pi}{2} \int \!\!\!\int dx dp \,W(x,p) \left( \frac{\partial^2}{\partial x^2} + \frac{\partial^2}{\partial p^2} + 2\right) W(x,p).
\label{eq: macroscopicity def}
\end{equation}
The macroscopicity is related to the sharpness of the oscillations of the state in phase space, and therefore directly linked with the separation of the macroscopic constituents of the state. For a pure mechanical state, the measure is identical to the number of phonons associated with the quantum fluctuations, which in turn is related to the quadrature variances: $\langle n\rangle= (V_{x_M} + V_{p_M})/2-1/2$~\cite{Laghaout2015}. From this simple relation it is clear that both the squeezing operation as well as the photon subtraction operations will increase the macroscopicity of the optical state. However, since the mechanical state is not in a pure state due to the finite optomechanical coupling strength and the finite temperature of the oscillator, the simple relation does not hold and the induced mechanical macroscopicity will be reduced.

\begin{figure}[tbp]
\includegraphics[width=1\columnwidth]{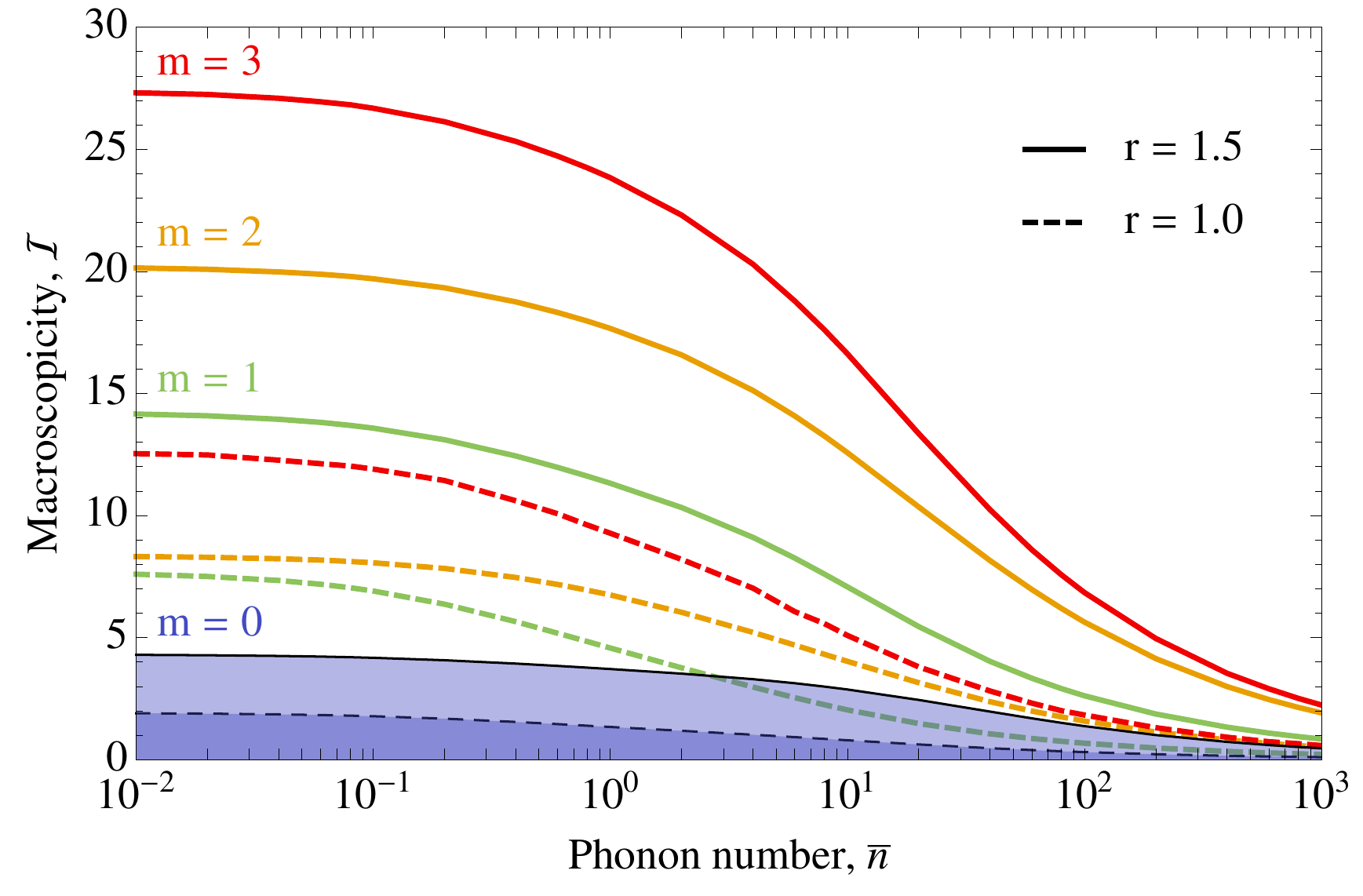}
\caption{Macroscopicity of the prepared mechanical state as a function of initial mechanical mode phonon occupancy, input squeezing $r$, and number of subtracted photons $m$. The pulse photon number is fixed to $N_p = 2.3 \cdot 10^5$, yielding an interaction strength $\chi=1$. Displacement amplitude fluctuations are assumed negligible. The shaded regions corresponding to $m=0$ indicate the contribution of optical squeezing alone to the total mechanical state macroscopicity.
\label{fig: macroscopicity} 
}
\end{figure}

Combining Eqs.~(\ref{eq: macroscopicity def}) and (\ref{eq: WmechOut}) we evaluate the macroscopicity of the final mechanical state as function of initial thermal phonon occupation, employing PSSV input states with $m\leq 3$ and a fixed displacement-enhanced interaction strength $\chi =1$. As presented in Fig.~\ref{fig: macroscopicity}, we find that, despite the finite interaction strength, the mechanical oscillator can be prepared in a state with appreciable macroscopicity even for $\bar{n}$ orders of magnitude above the ground state, increasing with $r$ and $m$. Taking $\mathcal{I}=4$ as a benchmark for truly macroscopic states, corresponding to a pure cat state with two shot noise units separation of the constituent coherent components~\cite{Jeong2005}, we find that this is achieved using a PSSV with $r=1.5$ and $m=3$ and an initial phonon occupation of $\bar{n}<200$. For $\bar{n}<1$ the macroscopicity increases to $\mathcal{I}>24$. Relaxing the optical input state to $r=1$ and $m=1$ significantly increases the constraint on the mechanical pre-cooling, requiring $\bar{n}$ below 2 to reach the macroscopicity benchmark.

\begin{figure}[tbp]
\includegraphics[width=.9\columnwidth]{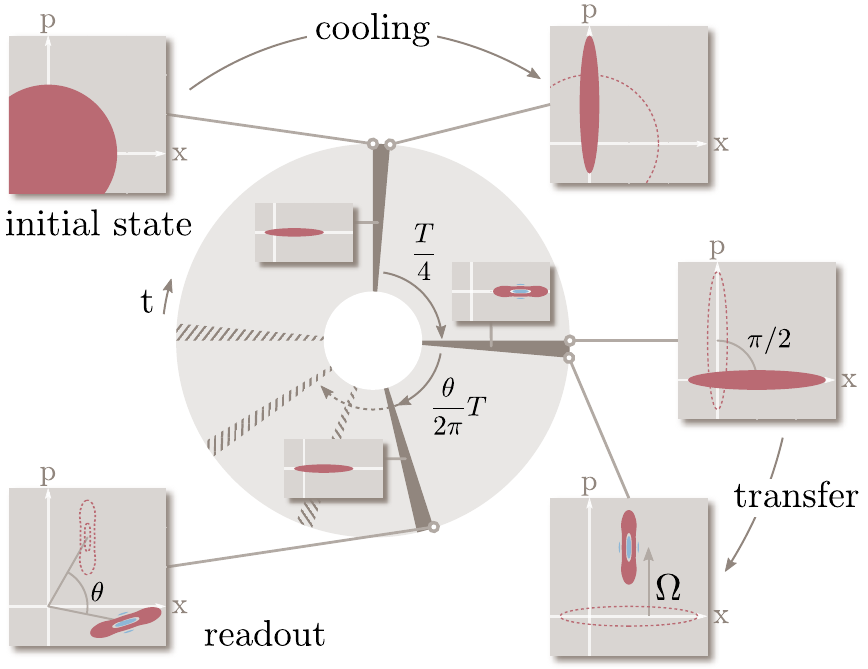}
\caption{The three-pulse protocol illustrated on a ``clock'' corresponding to one mechanical oscillation period, $T$. The first pulse of squeezed light (inner Wigner function) cools one quadrature of the initially thermal mechanical state (outer Wigner function). A quarter period later, the second pulse transfers the photon-subtracted state onto the mechanics. Finally, after a variable interval the third pulse reads out a rotated quadrature of the state.
All three QND interactions are accompanied by a homodyne detection of the reflected optical field (not shown).
\label{fig: 3pulse} 
}
\end{figure}
\paragraph{Three-pulse protocol.}
Up to now, we have considered the mechanical oscillator to be initialized in a phase space-symmetric thermal state, prior to the non-Gaussian state preparation, and tacitly assumed that the final oscillator state can be read out perfectly. However, as can be seen directly from Eqs.~(\ref{eq: xLout})-(\ref{eq: pMout}) the measurement-induced QND mapping of the optical amplitude quadrature onto the mechanical position variable is independent of the initial noise on the mechanical position. The state preparation process is only affected by the noise of the mechanical momentum variable and it is therefore sufficient to ``cool'' only that motional degree of freedom. This can be achieved by introducing a squeezed light QND interaction and allowing a subsequent quarter period harmonic evolution of the mechanical oscillator, immediately preceding the actual state preparation interaction.
Similarly, a squeezed light QND interaction can be used for optical readout of the mechanical oscillator, succeeding the non-Gaussian state preparation stage. We therefore suggest a three-pulse protocol in which the mechanical oscillator is asymmetrically cooled, prepared in a macroscopic superposition state, and finally optically characterized. The complete cooling-preparation-readout scheme is illustrated in Fig.~\ref{fig: 3pulse}.

Vacuum squeezed light is produced continuously while the displacement operation is pulsed and synchronized with the mechanical oscillator. 
The first and the third pulse comprise displaced phase squeezed vacuum states while the second pulse is a displaced PSSV state. With the first pulse, we use the QND interaction followed by homodyne detection to prepare the harmonic oscillator in a thermally squeezed state -- a process that is enhanced due to the squeezing of the input optical state. As discussed, this corresponds to asymmetric cooling where the oscillator noise is deamplified in a single variable, $x_M$, and amplified in the conjugate variable, $p_M$. A quarter period later the reduced noise is transferred to the momentum quadrature, cueing the second and principal light-mechanics interaction which effectuates the QND state preparation scheme discussed above. Finally, the last pulse is used to read out the mechanical quadratures, again using a QND coupling followed by homodyne detection. For full state tomography, the delay between the second and third pulse is varied in order to map out a range of quadrature phases within a half mechanical period.
The resolution of this readout measurement is given by the interaction strength $\chi$ and the noise of the optical pulse. It is thus important to note that due to the squeezing of the optical quadrature, the resolution of the measurement is greatly improved compared to the coherent state based protocol proposed in Ref.~\cite{Vanner2011}. Cooling and readout processes using a QND coupling and coherent states of light have already been experimentally demonstrated~\cite{Vanner2013}. 

To investigate further the feasibility of measurement-induced cooling, we deduce the resulting asymmetrically cooled mechanical state quadrature variances from Eq.~(\ref{eq: WmechOut}). Considering the case where only displaced phase squeezed vacuum is employed (no photon subtraction), we find
\begin{align}
V_{x_M}^{c} &= \frac{V_{x_M}}{1+\chi_c^2 V_{x_M}/V_{p_L}} \label{eq: precool_xm}\\
V_{p_M}^{c} &= V_{p_M} + \chi_c^2 V_{x_L} \label{eq: precool_pm},
\end{align}
where $\chi_c$ is the pre-cooling interaction strength. The single-quadrature cooling effect and the benefit of using squeezed light is immediately evident from the above expressions. Assuming a mechanical oscillator initially in equilibrium with a thermal bath at a temperature of $T_{bath}= 100\,\textrm{mK}$, corresponding to a phonon occupancy of $2.1 \cdot 10^6$, and using Eqs.~(\ref{eq: precool_xm}) and (\ref{eq: precool_pm}) to evaluate the pre-cooled mechanical quadrature variances, we have studied the total negativity of a non-Gaussian mechanical state prepared by the second QND interaction, as a function of the pre-cooling strength. Using a strongly squeezed cooling pulse, only a weak interaction strength of about $\chi_c = 0.2$, independent of displacement amplitude noise, is required to saturate the total negativity. This corresponds to a cooling pulse photon number on the order of $10^4$. 

Though sufficient for the state preparation alone, the proposed dynamical “cooling” protocol should be accompanied by standard passive cooling to mitigate the process of thermal decoherence. 
This process must not be significant during a single period of the mechanical oscillator to allow for cooling, preparation, and readout before coupling to the environment's thermal noise perturbs the quantum state of the system. Meeting this condition requires both the thermal heating of the pre-cooled mechanical state, occuring at a rate proportional to $\bar{n}_{th}\Gamma_M$, with $\bar{n}_{th}$ being the mean phonon number of the environment and $\Gamma_M$ the mechanical damping rate, as well as the thermal decoherence of the prepared macroscopic state to be much slower than a mechanical oscillation period. For a mechanical cat state with a phase space separation of the coherent amplitudes of $d=10$, similar to the $m=3$ state in Fig.~\ref{fig: CondmechWignerstates}, the decoherence time is $\tau_{dec} = (\bar{n}_{th}\Gamma_M (d\sqrt{2})^2/6)^{-1} = 230\,\mu {\rm s}$~\cite{OConnell2003}, and for the system in question, heating out of the motional ground state happens on a time scale of $\tau_{th} = 7.6\,{\rm ms}$. Comparing this to the mechanical oscillation period $T= 10\,\mu{\rm s}$ we see that the proposed protocol is indeed a viable approach to demonstration of truly macroscopic quantum states of mechanical motion.     
 
\paragraph{Conclusion.} 
We have presented a protocol for generation of Schr\"odinger cat-like states of a macroscopic mechanical oscillator, relying on previously demonstrated techniques and compatible with existing cavity optomechanical systems. By taking advantage of squeezed-light enhanced quantum non-demolition interactions, non-Gaussian resources, and homodyne detection we have circumvented the demanding requirements of strong single-photon coupling and operation in the sideband resolved regime. Consequently, our results pave a feasible route towards the long-standing goal of interrogating quantum mechanical phenomena at the macroscopic scale.

\paragraph{}
This work was supported by the Danish Council for Independent Research (Sapere Aude program), the Lundbeck Foundation, and the Villum Foundation (Young Investigator Programme). We thank Anders S. Sørensen for helpful discussions.

\bibliography{macromech_bibliography}

\end{document}